\documentclass[aps,prc,nofootinbib,showpacs]{revtex4}

\usepackage{epsfig}
\usepackage{graphicx}

\begin{document}
\title{Two-component model for the deuteron electromagnetic structure}

\author{E. Tomasi-Gustafsson}
\affiliation{\it DAPNIA/SPhN, CEA/Saclay, 91191 Gif-sur-Yvette Cedex, 
France}
\author{G. I. Gakh}
\altaffiliation{Permanent address:
\it NSC Kharkov Physical Technical Institute, 61108 Kharkov, Ukraine}

\affiliation{\it DAPNIA/SPhN, CEA/Saclay, 91191 Gif-sur-Yvette Cedex, 
France}

\author{C. Adamu\v s\v c\'in}
\altaffiliation{Leave on absence from
\it Department of Theoretical Physics, IOP, Slovak Academy of Sciences, Bratislava, Slovakia}

\affiliation{\it DAPNIA/SPhN, CEA/Saclay, 91191 Gif-sur-Yvette Cedex, 
France}

\date{\today}

\pacs{}
\def\ges{$G_{Es}$}

\def\gep{$G_{Ep}$}
\def\gen{$G_{En}$}
\def\vqb{{\vec {q_B}}}

\begin{abstract}
We suggest a simple phenomenological parametrization for all three deuteron electromagnetic form factors, and show that a good fit on the available data,  with a minimal number of parameters, can be obtained. The present description of the deuteron electromagnetic structure is based on two components with different radii, one corresponding to two nucleons separated by $\simeq $2 fm, and a standard isoscalar contribution, saturated by $\omega$ and $\phi$ mesons, only. 
\end{abstract}
\maketitle
\section{Introduction}

The electromagnetic structure of hadrons is characterized by a set of electromagnetic form factors (FFs), which are functions of one variable, the four momentum transfer squared, $Q^2$.

Since the pioneering work of Hofstadter \cite{Ho62}, a large number of experimental data have been collected about hadron electromagnetic FFs, especially in the space-like region. It is important to note that a relatively simple phenomenological parametrizations can be found for the description of the $Q^2$ dependence of different FFs, despite the possible complicated dynamics which determines the hadronic electromagnetic structure. Therefore, even without a complete understanding of the internal structure of the hadrons, simple analytical formulas have been suggested for the hadronic FFs.

For example, in case of charged pion, the corresponding FF, $F_{\pi}(Q^2)$, can be written as:
\begin{equation}
F_{\pi}(Q^2)=(1+Q^2/m^2_{\rho})^{-1},
\label{eq:eq1a}
\end{equation}where $m_{\rho}$ is the $\rho$-meson mass, and $Q^2\ge 0$ in the space-like region.

Similar formulas have also been suggested for FFs of the electromagnetic transition $\gamma+\gamma^*\to P^0$, $P^0=\pi^0$, $\eta$, $\eta'$ and $\eta_c$:
\begin{equation}
F_{P\gamma\gamma^*}(Q^2)=F_P(0)(1+\gamma Q^2/m_P^2)^{-1},
\label{eq:eq2a}
\end{equation}
where $m_P $ is a fitting parameter, which depends on the type of pseudoscalar meson.
  
For a long time, the nucleon electromagnetic FFs have been described by a very simple form:
\begin{equation}
G_{Ep}(Q^2)=G_{Mp}(Q^2)/\mu_p= G_{Mn}(Q^2)/\mu_n=
[1+Q^2~[\mbox{GeV}^2]/ 0.71 ]^{-2},~G_{En}(Q^2)=0,
\label{eq:eq3}
\end{equation}
where $\mu_p=2.79(\mu_n=-1.91)$ is the magnetic moment of proton (neutron).  
But the last experiments at the Jefferson laboratory, using the polarization transfer method \cite{Re68}, showed a large deviation from parametrization (\ref{eq:eq3}) and the proton electromagnetic FFs have, instead, the following  behavior  \cite{JGP00}:
\begin{equation}
\mu_pG_{Ep}(Q^2)/G_{Mp}(Q^2)=1-0.13(Q^2 \mbox{~[GeV]}^2-0.04).
\label{eq:eq4}
\end{equation}
The experimental data about $G_{En}(Q^2)$, obtained from quasielastic scattering of longitudinally polarized electrons by a polarized deuteron target, $\vec d(\vec e, en)p$ \cite{Wa04}, and from the measurement of the neutron polarization in $ d(\vec e, e'\vec n) p$ \cite{Ma03}, show, indeed, that $G_{En}(Q^2)\ne 0$, with the following parametrization \cite{Ga71}: 
$G_{En}(Q^2)=-\eta G_{Mn}/(1+5.6 \eta) $, with $\eta=Q^2/(4m^2)$, and $m$ is the nucleon mass.

Note that the analysis \cite{ETG02} of all existing experimental data concerning elastic electron-deuteron $(ed)$ scattering - the structure functions $A(Q^2)$ and $B(Q^2)$ and the deuteron tensor polarization - in framework of the impulse approximation, leads to comparable values of $G_{En}(Q^2)$ and $G_{Ep}(Q^2)$ at relatively large momentum transfer, $Q^2\ge 2$ GeV$^2$.

Next in the list of simple parametrizations of hadronic electromagnetic FFs, we can  mention that the magnetic FF for the transition $N\to \Delta (1232)$, ${\cal J}^P=3/2^+$ can also be parametrized in a simple form \cite{Fr99}:
\begin{equation}
G_{N\Delta}(Q^2)=G_{N\Delta}(0)(1+Q^2/m^{*2})^{-2},~ m^{*2}\le 0.71 \mbox{~GeV$^2$}.
\label{eq:eq5}
\end{equation}
Finally, the nucleon axial FF, $G_A(Q^2)$, for the transition $W^*+p\to n$ ($W^*$ is the virtual $W$-boson), can be described by the following simple formula \cite{Be02}:
 \begin{equation}
 G_A(Q^2)=G_A(0)(1+Q^2/m_A^2)^{-n}
\label{eq:eq6}
\end{equation}
with $m_A$ = 1.06 GeV, if $n=2$.

These simple parametrizations are very useful for the discussion of different processes of elastic and inelastic scattering of electrons on nuclei at high energies. Nevertheless, these parametrizations can not be considered a complete and precise description of the hadronic electromagnetic structure in the full region of momentum transfer (space-like and time-like region as well). An evident example is that the parametrizations (\ref{eq:eq3}) and (\ref{eq:eq4}) suggested for the description of the nucleon structure in the space-like region, violate the relation
$G_{En}(Q^2)= G_{Mn}(Q^2)$, at $ Q^2=-4m^2$,
i.e., at the threshold of the annihilation process $e^++e^-\to N+\overline{N}$.

To avoid this and other problems, in Ref. \cite{Ia73} (updated in Ref. \cite{Bi04}),  another parametrization of nucleonic FFs has been suggested, which can be extended in the whole region of momentum transfer squared. The basic idea of this parametrization is the presence of two components in the nucleon structure, with different radii: the intrinsic structure, with radius $\simeq 0.34$ fm (updated 0.49 fm), characterized by a dipole FF (which is the same for electric, magnetic, proton and neutron FFs) and a meson cloud, (which contains the $\rho$, $\omega$, $\phi$ contributions)  different for each of the four nucleon FFs.

In this paper we generalize this two-component picture to the case of deuteron electromagnetic FFs, with the aim to find a simple parametrization for all three FFs, at least in the region where they have been completely determined, for $Q^2\le $ 2 GeV$^2$. This parametrization should be useful in corresponding calculations of deuteron electromagnetic processes, such as $e^-+d\to e^-+d$,  $e^-+d\to e^-+n+p$, $e^-+d\to e^-+d+\pi^0$ etc.

Any specific additional assumption about the validity of impulse approximation, the role and size of meson exchange currents, relativistic corrections, model (relativistic or non relativistic) of deuteron structure etc. is not needed. Instead, we parametrize the three deuteron FFs in a simple form, with a small number of parameters, normalized for $Q^2=0$ to the electric charge, the magnetic moment and the electric quadrupole moment of the deuteron. In order to decrease the number of independent parameters, we will use the experimental constraint of the position of the node for the electric and magnetic FFs.

\section{Formalism}

\begin{figure}
\mbox{\epsfxsize=14.cm\leavevmode \epsffile{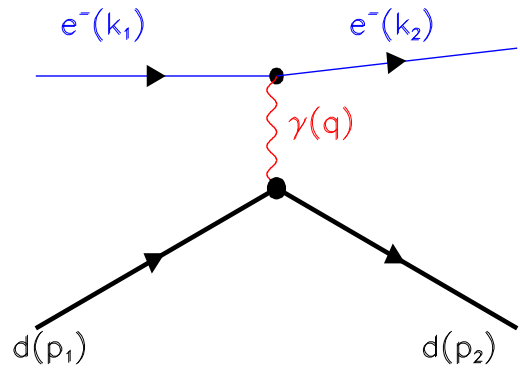}}
\vspace*{.2 truecm}
\caption{Feynman diagram for electron-deuteron elastic scattering, within the one-photon mechanism.}
\label{fig:fig1}
\end{figure}

The matrix element for $ed$ elastic scattering, for the one-photon approximation, Fig. \ref{fig:fig1}, can be written as:
\begin{equation}
{\cal M}=\displaystyle\frac{e^2}{Q^2}\overline{u}(k_2)\gamma_{\mu}u(k_1) {\cal J}_{\mu}(d), 
\label{eq:med}
\end{equation}
where $k_1$ and $k_2$ are the four momenta of the initial and final electron, and ${\cal J}_{\mu}(d)$ is the electromagnetic current of the deuteron. Applying the conservation of this current, the P and C invariance of the electromagnetic interaction of hadrons, we can write ${\cal J}_{\mu}(d)$ in the following general form \cite{Ca61}:
\begin{eqnarray}
{\cal J}_{\mu}(d)&=- & \left \{G_1(Q^2)(U_2^*\cdot U_1)(p_1+p_2)_{\mu}+  ~~G_2(Q^2)[U_{1\mu}(U_2^*\cdot q) -U_{2\mu}^*(U_1\cdot q)]- \right .\label {eq:cal}\\
&& ~~\left . G_3(Q^2)\displaystyle\frac{1}{2M^2}(U_1\cdot q)
(U_2^*\cdot q)(p_1+p_2)_{\mu} \right \},\nonumber
\end{eqnarray}
where $p_1(U_1)$ and $p_2(U_2)$ are the four momenta (polarization four vectors) of the initial and final deuteron, and the polarization four vectors satisfy the condition: $p_1\cdot U_1= p_2\cdot U_2=0$, $M$ is the deuteron mass.

The deuteron invariant form factors $G_i$, $i=1-3$, are related to the charge, quadrupole and magnetic FFs by: 
\begin{equation}
G_c=G_1+\displaystyle\frac{2}{3}\tau G_2,~
G_m=G_2, ~
G_q=G_1-G_2+(1+\tau) G_3,~\tau =\displaystyle\frac{Q^2}{4M^2}.
\label{eq:eqff}
\end{equation}

The differential cross section for elastic electron deuteron scattering can be expressed in terms of two structure functions, $A(Q^2)$  and $B(Q^2)$, which  depend on the three electromagnetic FFs:
\begin{equation} 
{{d\sigma} \over {d\Omega}} = \sigma_M \left[ A(Q^2) +
B(Q^2) \tan^2{\left(\frac{\theta}{2}\right)} \right ] ,  
\label{eq:eq10}
\end{equation}
where $\sigma_M=\alpha^2 E^\prime \cos^2(\theta/2)/[4 E^3 \sin^4(\theta/2)]$
is the Mott cross section.  
Here $E$ and $E^\prime$ are the incident and scattered electron
energies, $\theta$ is the electron scattering angle,
$Q^2=4 E E^\prime \sin^2(\theta/2)$ is the four-momentum transfer
squared and $\alpha$ is the fine structure constant, $\alpha=e^2/4\pi=1/137$.  The elastic
electric and magnetic structure functions $A(Q^2)$ and $B(Q^2)$ are
written in terms of the charge, quadrupole and magnetic FFs $G_c(Q^2)$, $G_q(Q^2)$, and $G_m(Q^2)$ as:
\begin{equation}
{ A(Q^2) =  { G^2_c(Q^2) + {8 \over 9} \tau^2 G^2_q(Q^2) +
{2 \over 3} \tau G^2_m(Q^2) }  },~
{ B(Q^2) =  {4 \over 3} \tau (1+\tau) G^2_m(Q^2) }.
\label{eq:eq11}
\end{equation}
In order to determine the three FFs, one needs another observables, usually the component $t_{20}$ of the tensor polarization of the recoil deuteron, in an unpolarized collision, which contains the following combination of the three FFs:
$$
t_{20}=-\displaystyle\frac{1}{\sqrt{2}{\cal S}}\left \{
\displaystyle\frac{8}{3}\tau G_c G_q + \displaystyle\frac{8}{9} \tau^2 
G_q^2+\frac{1}{3} \tau
\left [1+2(1+\tau)\tan^2 (\theta /2)\right ]G_m^2\right \},
$$
where ${\cal S}=A+B\tan^2 (\theta /2)$. The existing data on the differential cross section \cite{Al99} and $t_{20}$ \cite{Ab00} for electron deuteron elastic scattering allow the extraction of the three electromagnetic deuteron FFs up to $Q^2\simeq $ 2 GeV$^2$. This has been done in Ref. \cite{www} where the world data were collected and three different analytical parametrizations were suggested, with a number of parameters varying from 12 to 33. 

In general, the existing  parametrizations of deuteron electromagnetic FFs contain a large number of parameters, and are often based on analytical formulas, with poor physical content. An attempt to find a global description based on the vector dominance model, and  satisfying the asymptotic conditions predicted by QCD at large $Q^2$, \cite{Ko95}, lead to a twelve parameters fit. The fit was updated including the  world data on $ed$ elastic scattering in Ref. \cite{www} where two other fits were suggested. One is a sum of inverse polynomial terms, where the first node of the corresponding FFs was introduced in a global multiplicative term. The number of free parameters,  necessary to obtain $\chi^2/ndf=1.5 $, was eighteen. The last  parametrization is a sum of gaussians, with some physical constraints on the parameters, which are the width and the position of the maximum of the gaussians. In total the parametrization contains 33 parameters for $\chi^2/ndf=1.5 $.

We suggest here a simpler parametrization, based on transparent physical content, with a minimal number of parameters. More precisely, we extend a  two-component parametrization, already successfully applied to nucleon electromagnetic FFs \cite{Ia73,Bi04} and recently to strange nucleon FFs \cite{Ia73,Bi05}, to the deuteron electromagnetic FFs.

The deuteron is an isoscalar particle, therefore, considering only the contribution of the isoscalar vector mesons, $\omega$ and $\phi$, one can write:
\begin{equation}
G_i(Q^2)=N_i g_i(Q^2) F_i(Q^2),~i=c,q,m
\label{eq:eq1}
\end{equation}
with: 
$$F_i(Q^2)= 1-\alpha_i-\beta_i+
\alpha_i\displaystyle\frac{m_{\omega}^2}{m_{\omega}^2+Q^2}
+\beta_i\displaystyle\frac{m_{\phi}^2}{m_{\phi}^2+Q^2}. $$
where $m_{\omega}$ ($m_{\phi}$) is the mass of the $\omega$ ($\phi$)-meson. Note that the $Q^2$ dependence of $F_i(Q^2)$ is parametrized in such form that $F_i(0)=1$, for any values of the free parameters $\alpha_i$ and $\beta_i$, which are real numbers.

The terms $g_i(Q^2)$ are written as functions of two  parameters, also real, $\gamma_i$ and $\delta_i$, generally different for each FF:
\begin{equation}
 g_i(Q^2)=1/\left [1+\gamma_i{Q^2}\right ]^{\delta_i},
\label{eq:eq12}
\end{equation}
and $N_i$ is the normalization of the $i$-th FF at $Q^2=0$:
$$N_c=G_c(0)=1,$$
$$N_q=G_q(0)=M^2{\cal Q}_d=25.83,$$
$$N_m=G_m(0)=\displaystyle\frac{M}{m}\mu_d=1.714,$$
where ${\cal Q}_d$, and $\mu_d$ are the quadrupole and the magnetic moments of the deuteron.

\section{Results and discussion}

In Ref. \cite{www} the existing data on $ed$ elastic scattering, differential cross section and the polarization observables, were reconsidered. Values of the three deuteron FFs were extracted, for the $Q^2$ values where $t_{20}$ measurements were available, getting the values of $A(Q^2)$ and $B(Q^2)$ from an interpolation of the data on the differential cross section. While the magnetic  FF, $G_m(Q^2)$, is directly related to $B(Q^2)$, the extraction of the charge and quadrupole FFs requires the solution of two quadratic equations, which may lead, in some cases, to two possible roots. Therefore, the analysis of $G_c(Q^2)$ and $G_q(Q^2)$ consists in two different sets of solutions and two corresponding fits. 

The experimental data for $G_c$ and $G_m$ show the existence of a zero, for $Q_{0c}^2\simeq 0.7$ GeV$^2$ and $Q_{0m}^2\simeq 2$ GeV$^2$. The constraint of a node gives the following relation between the parameters $\alpha_i$ and $\beta_i$, $i=c$ and $m$:
\begin{equation}
\alpha_i=\displaystyle\frac{m_{\omega}^2+Q_{0i}^2}{Q_{0i}^2}-
\beta_i\displaystyle\frac{m_{\omega}^2+Q_{0i}^2}{m_{\phi}^2+Q_{0i}^2}.
\label{eq:eqal}
\end{equation}
In the fitting procedure this relation allows to obtain a better description of the data and a faster convergency, reducing the number of free parameters.

The expression (\ref{eq:eq1}) contains four parameters, $\alpha_i$, $\beta_i$, $\gamma_i$, $\delta_i$, generally different for different FFs. We consider the region $Q^2\le 2$ GeV$^2$, where the separation of $G_c$ and $G_q$ has been done.

In principle the parameters $\gamma$ and $\delta$, Eq. (\ref{eq:eq12}), may be fixed by the asymptotic behavior of the deuteron FFs, which follows from quark counting rules \cite{Ma73}. However, the range of applicability of the present parametrization is {\it a priori} restricted to $Q^2\le 2$ GeV$^2$ and this region is expected to be far from the asymptotic region \cite{Re03}. 

The data basis of the present study consists in the data tabulated as in Table I of Ref. \cite{www} \footnote{The value for $G_q$ corresponding to $Q=2.788$ fm$^{-1}$ should be $2.59^{+0.07}_{-0.71}$, instead of $2.59(\pm 0.073)$ \protect\cite{Ba06}} and completed by more recent measurements from Ref. \cite{Ni03}.

As a result of the procedure for the extraction of the values of $G_{c}(Q^2)$ and
$G_{q}(Q^2)$ from $A(Q^2)$, $B(Q^2)$ and $t_{20}(Q^2)$, some experimental points show  a large asymmetry of the errors, which can not be neglected in this  analysis. While there is still no general guide how to treat asymmetric errors, we used two
different ways to handle them. At first no asymmetry was assumed and the  average of the upper ($\sigma_+$) and lower ($\sigma_-$) errors was taken. 
Then, two approaches recently proposed in Ref. \cite{Ba04} were applied.  Model 1, as proposed in the paper, assumes linear dependencies and defines the contribution to a modified $\chi^{2}$ as
\begin{equation}
\chi ^2=\sum_i\frac{\epsilon_i^2}{\sigma_{-i}^2},\mbox{~~~for~~}\epsilon_i>0,~
\chi^2=\sum_i\frac{\epsilon_i^2}{\sigma_{+i}^2},\mbox{~~~for~~}\epsilon_i<0,
\label{eq:eqfit}
\end{equation}
where $\epsilon_i$ is the discrepancy between the $i$-th experimental point and the  value of the corresponding function. 
Model 2, preferred by the author, is based on a quadratic approximation. It 
was not always suitable for our analysis, because, in some cases it doesn't give real solutions for the contribution to $\chi^{2}$. 

In any case, the analysis which takes into account the asymmetry of errors 
(\ref {eq:eqfit}) gives significant reduction of $\chi^{2}$ in all cases, but it didn't influence significantly the resulting parameters of the fit, except for $G_{q}(Q^2)$, where the errors on the parameters were significantly reduced.

The results were firstly obtained with a three parameter fit
$\beta$, $\gamma$, $\delta$, and the constraint (\ref{eq:eqal}) for $G_c$  and $G_m$ and a four parameter fit $\alpha$, $\beta$, $\gamma$, $\delta$, according to Eq.  (\ref{eq:eq1}), for $G_q$.

The parameters $\delta$ and $\gamma$ are similar for all FFs $G_c$ and $G_m$, with good accuracy (Table \ref{tab1}). A change within 10\% in the position of the node, slightly  affects the quality of the fit, improving in general one parametrization, while the other gets worse. The fit is quite sensitive to the choice of initial parameters, in particular for $G_q$. In case of $G_q$, which is not constrained by a node, a good fit can be obtained with a large cancellation of the terms driven by $\alpha$ and $\beta$.

The parameters $\gamma$ and $\delta$ characterize the global structure of the deuteron, and the factor $1/(1+\gamma Q^2)^{\delta}$ is related to the two nucleon core of the deuteron. But the isoscalar structure of the electromagnetic structure of the deuteron FFs, which is described by the functions $F_i(Q^2)$, in terms of $\phi$- and $\omega$-mesons contributions, is  different for the three FFs, with different sets of parameters $\alpha$ and $\beta$. 

From Table \ref{tab1} one can see that the parameters $\gamma$ and $\delta$ are not so different for the three FFs. This means that FFs would be mostly sensitive to the meson cloud. In order to test this, a global fit was performed, keeping the $\gamma$ and $\delta$ the same for the three FFs, and fitting (or fixing) $\alpha$ and $\beta$ as previously. In such fit, two solutions appear also for $G_m$, related to the choice of the other two FFs.

\begin{figure}
\mbox{\epsfxsize=14.cm\leavevmode \epsffile{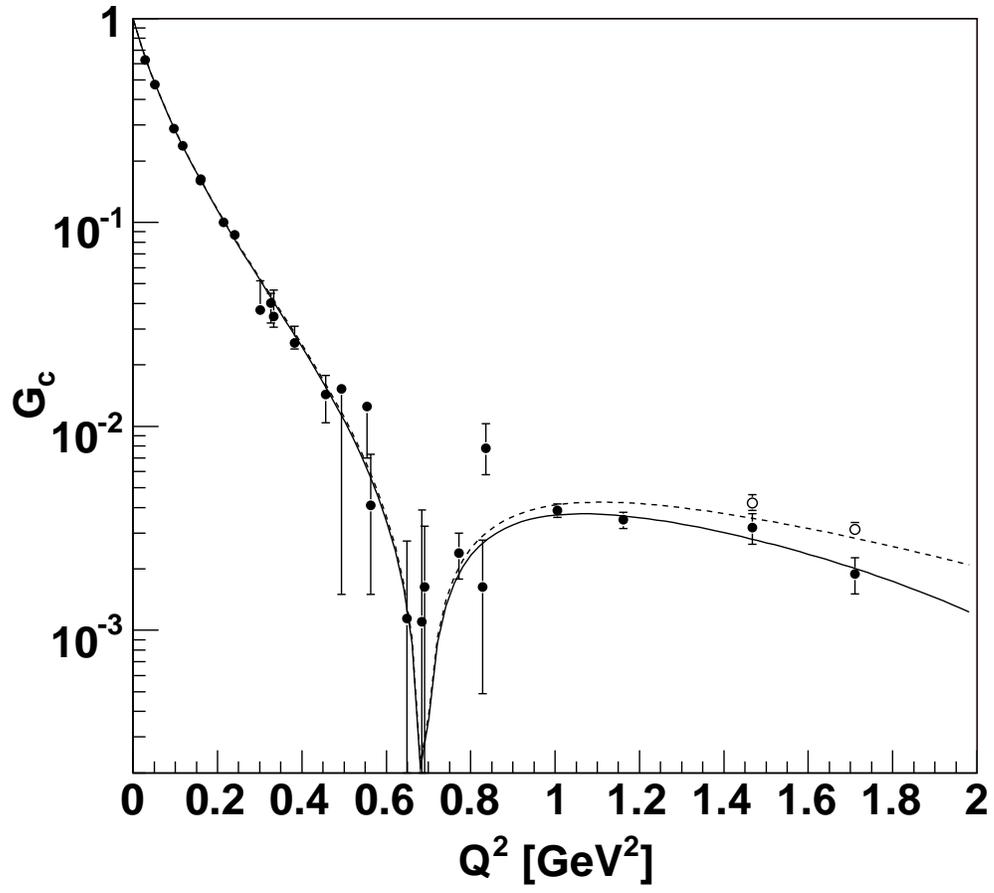}}
\vspace*{.2 truecm}
\caption{Fit to deuteron charge form factor data. The solid and dashed lines correspond to the fits for the two different solutions for the data (solid and empty circles).}
\label{fig:gc}
\end{figure}

\begin{figure}
\mbox{\epsfxsize=14.cm\leavevmode \epsffile{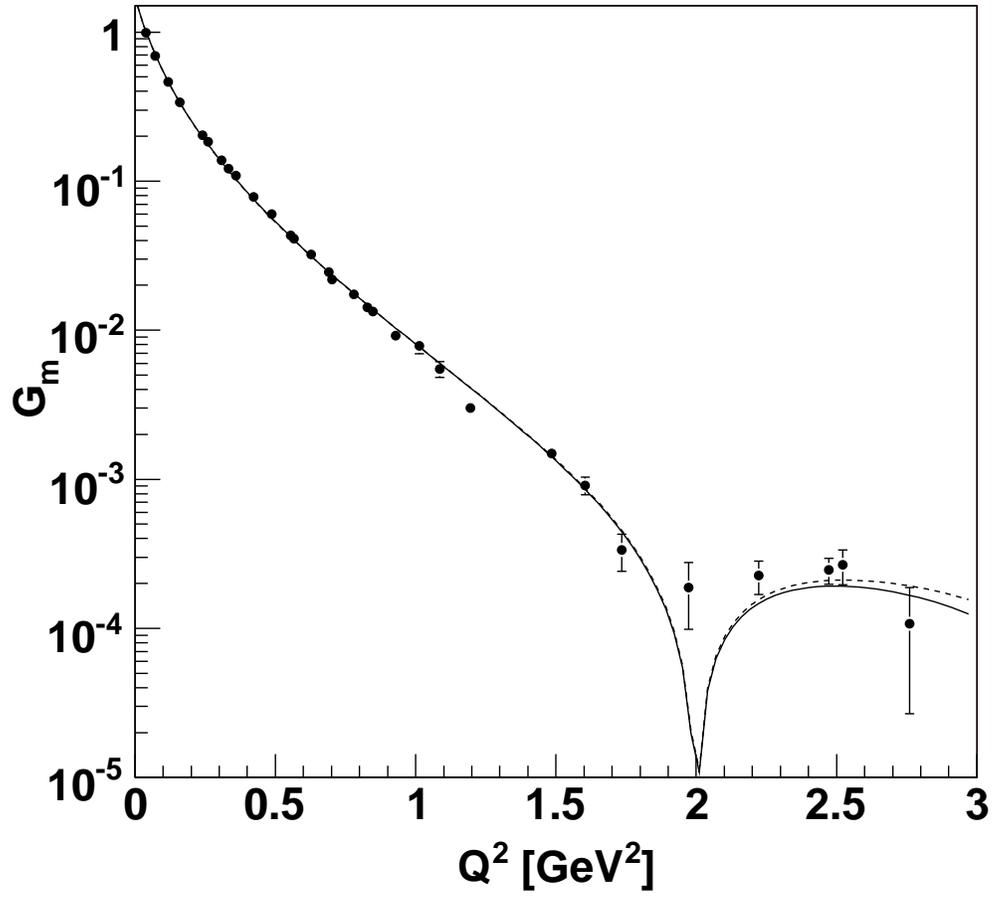}}
\vspace*{.2 truecm}
\caption{Fit to deuteron magnetic form factor data.}
\label{fig:gm}
\end{figure}

\begin{figure}
\mbox{\epsfxsize=14.cm\leavevmode \epsffile{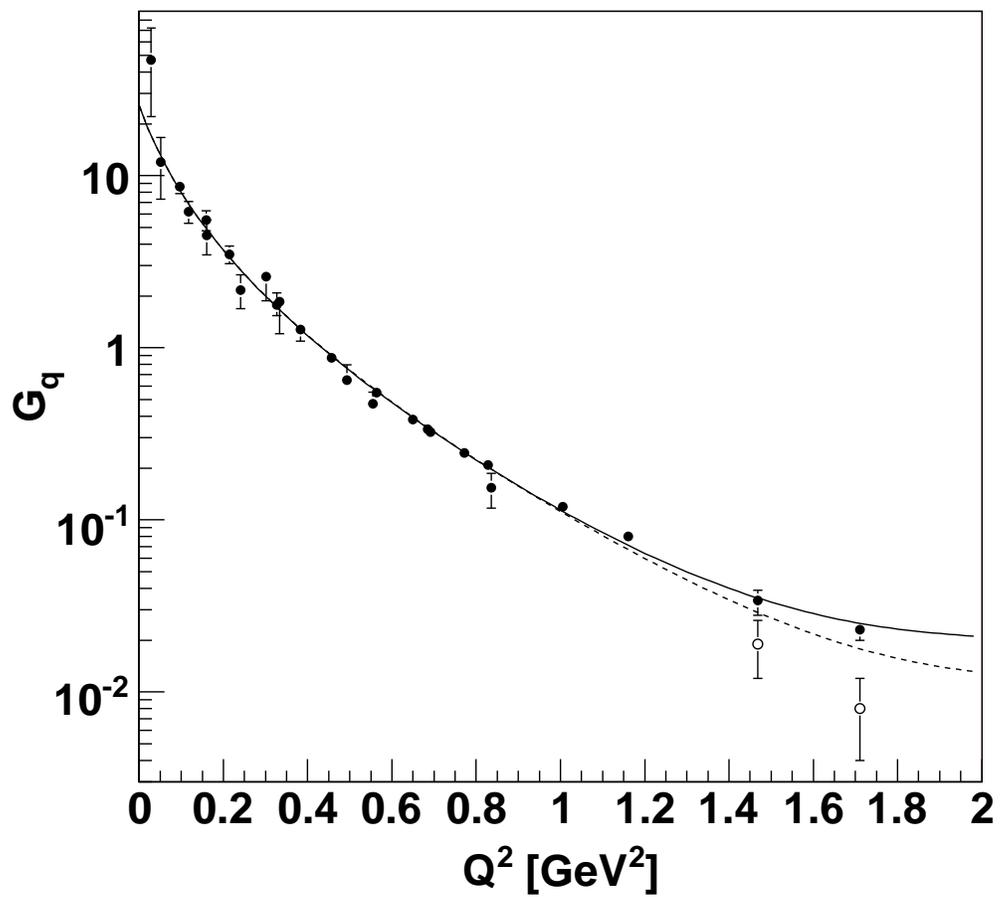}}
\vspace*{.2 truecm}
\caption{Fit to deuteron quadrupole form factor data. Notations as in Fig. \ref{fig:gc}.}
\label{fig:gq}
\end{figure}

In Figs. 
\ref{fig:gc}, \ref{fig:gm} and \ref{fig:gq}, the data points used in the present fit are shown, together with the result of this last fit (solid (dashed) lines correspond respectively to the first (second) parametrization). Open symbols in Figs. \ref{fig:gc} and \ref{fig:gq} correspond to the second solution for $G_c$ and $G_q$. The values of the best fit parameters are reported in Table \ref{tab2}. The common parameters are  $\delta=1.04\pm 0.03$, $\gamma=12.1\pm 0.5$, for the first solution, corresponding to $\chi2/ndf=1.1$, whereas, for the second one, $\delta=1.05\pm 0.03$, $\gamma=12.1\pm 0.5$ and $\chi2/ndf=1.5$.

\begin{table*}
\begin{tabular}{|c|c|c|c|c|c|}
\hline\hline
& $\alpha$ & $\beta$   & $\gamma$ [GeV]$^{-2}$&  $\delta$& $\chi2/ndf$ \\
\hline\hline
$G_c$ (I) &$5.9\pm 0.1$ & $-5.2\pm 0.2$ & $13.9 \pm 1.4$&$0.96 \pm 0.07$&  0.8\\
$G_c$ (II)  &$5.0\pm 0.2$ & $-4.5 \pm  0.3$ & $11.5 \pm  1.2$&$1.11 \pm 0.09$& 
1.2\\
\hline
$G_q$(I)&$3.1\pm 1.1$ & $-2.1\pm 1.2 $ & $7.2\pm 2.8$& $1.6\pm 0.5$ & 0.5\\
$G_q$(II)&$1.4\pm 2.0$ & $-0.1\pm 2.4 $ & $7.7\pm 1.6$& $1.7\pm 0.4$ & 0.8\\
\hline
$G_m$ & $3.78\pm 0.04$& $-2.87\pm 0.04 $ & $11.4 \pm 0.5$& $1.07\pm 0.03$ & 1.5 \\
\hline\hline
\end{tabular}
\caption{ Parameters for the three deuteron electromagnetic FFs. In case of
$G_c$ and $G_m$, $\alpha$ is not a parameter, but it is derived from Eq.
(\protect\ref{eq:eqal}).}
\label{tab1}
\end{table*}

\begin{table*}
\begin{tabular}{|c|c|c|}
\hline\hline
& $\alpha$ & $\beta$ \\
\hline\hline
$G_c$ (I) &$5.75\pm 0.07$ & $-5.11\pm 0.09$\\
$G_c$ (II)  &$5.50\pm 0.06$ & $-4.78 \pm 0.08$\\
\hline
$G_q$(I)&$4.21\pm 0.05$ & $-3.41\pm 0.07$\\
$G_q$(II)&$4.08\pm 0.07$ & $-3.25\pm 0.09$\\
\hline
$G_m(I)$ & $3.77\pm 0.04$& $-2.86\pm 0.05$\\
$G_m(II)$ & $3.74\pm 0.04$& $-2.83\pm 0.05$\\
\hline\hline
\end{tabular}
\caption{ Parameters $\alpha$ and $\beta$  obtained from a global fit of the three deuteron electromagnetic
FFs. The parameters $\delta$ and $\gamma$ are the same
for all form factors and in case of $G_c$ and $G_m$, $\alpha$ is not a
parameter, but it is derived from Eq. (\protect\ref{eq:eqal}).}
\label{tab2}
\end{table*}

In Ref. \cite{Br76}, in order to study the behavior of deuteron FFs, in framework of QCD,  a reduced deuteron FF, $f_R(Q^2)$, was defined as:
\begin{equation}
f_R(Q^2)=\sqrt{A(Q^2)}\left (1+\displaystyle\frac{Q^2}{m_0^2}\right )/
F_N^2(\displaystyle\frac{Q^2}{4}),
\label{eq:eqbr}
\end{equation}
where $m_0^2=0.28$ GeV$^2$ and $F_N$ is a generalized nucleon electromagnetic FF\footnote{A discussion of the dependence of $f_R(Q^2)$ on the choice of $F_N$ can be found in Ref. \protect\cite{Re03}.}.

Comparing Eqs. (\ref{eq:eq1}) and  (\ref{eq:eqbr}), one can identify $g_i(Q^2)$ with the term $(1+Q^2/m_0^2)^{-1}$, then one expects $\delta=1$ and $\gamma=1/m_0^2$. From Table 1 it appears that $\delta\simeq 1$, but $\gamma$ is larger and corresponds to $m_0^2\simeq 0.1$ GeV$^2$. The product 
 $R^2=6\gamma\delta $ is related to the radius of the two-nucleon (2N) component, and one finds: $R_{2N}$=1.7 fm, for $m_0^2\simeq 0.08 $~GeV$^2$.
 
Note, in this respect, that the standard nonrelativistic description of the deuteron results in $R\simeq 1/\sqrt{2mE_D}\simeq 4$ fm, where $E_D$ is the deuteron binding energy.

\begin{figure}
\mbox{\epsfxsize=14.cm\leavevmode \epsffile{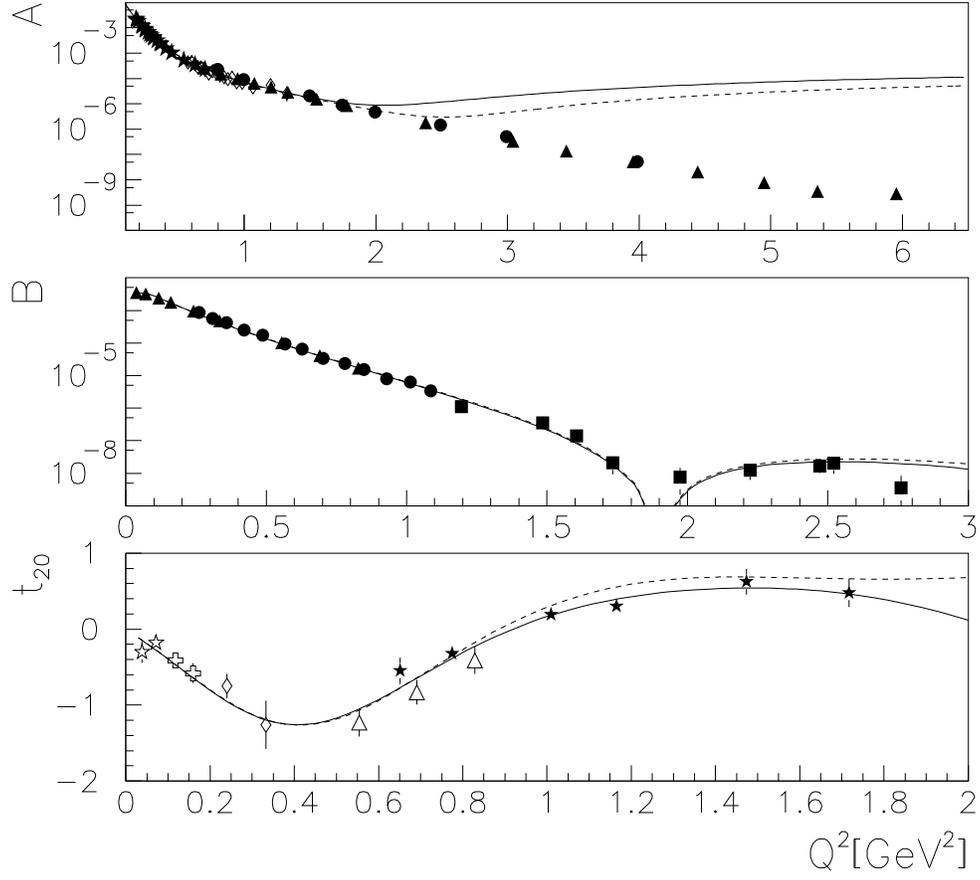}}
\vspace*{.2 truecm}
\caption{Illustration of the quality of the suggested fit on the experimental observables A, B and $t_{20}$, calculated with parametrization I (solid line) and II (dashed line).}
\label{fig:obser}
\end{figure}

\begin{figure}
\mbox{\epsfxsize=14.cm\leavevmode \epsffile{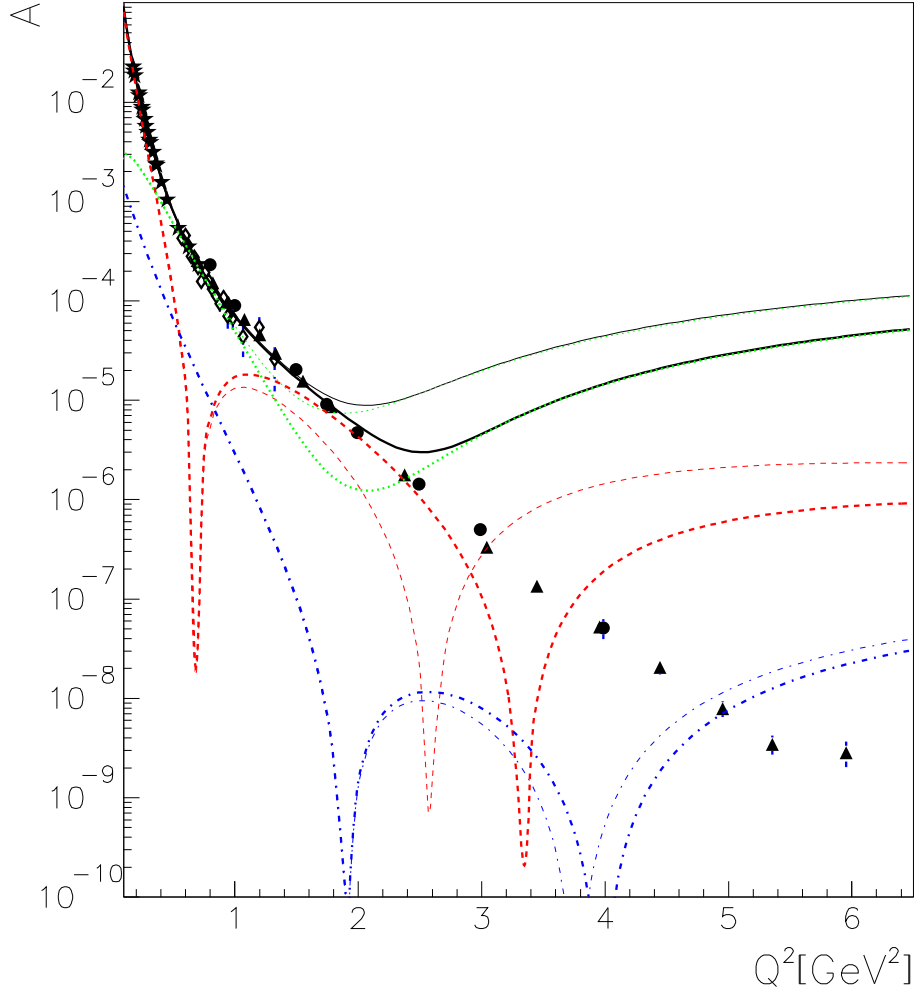}}
\vspace*{.2 truecm}
\caption{Different contributions to the structure function A, according to Parametrization I (thin lines)  and II (thick lines). The term related to $G_c$, $G_q$, and $G_m$, are shown as dashed, dotted, and dash-dotted lines, repectively.}
\label{fig:aq}
\end{figure}

Data are not expected to be extended at higher $Q^2$ in next future. So, the present parametrization can not be constrained at higher $Q^2$. Nevertheless, we compared the predictions for the structure function $A(Q^2)$, which has been measured up to 6 GeV$^2$, as well as for the observables $B(Q^2)$ and $t_{20}$ (Fig. \ref{fig:obser}). The description is good, as expected, in the range constrained by the fit. The difference between the experiment and the suggested parametrization visible for $A(Q^2)$ at larger $Q^2$, depends on the parametrization used, due in particular to the position of a second node which appears for $G_c(Q^2)$. The individual charge, magnetic and quadrupole  contributions to $A(Q^2)$ are shown in Fig. \ref{fig:aq}, for the two possible solutions. 

The discrepancy between the data for $A(Q^2)$ and the suggested deuteron FFs parametrizations can be interpreted as an indication that the region at  $Q^2\ge$ 2 GeV$^2$ is a transition region from the hadronic description of the deuteron structure to the quark degrees of freedom.

Another possible source of the difficulty to extend the fit at higher $Q^2$ may originate from the fact that at some point, at large $Q^2$, the one-photon approximation, that is at the basis of the relations used here among FFs and experimental observables, and usually assumed in electron hadron elastic and inelastic scattering,  does not hold anymore. A mechanism, where two photons, which equally share the momentum transfer squared, could become important. A discussion of this problem, concerning precisely $ed$ elastic scattering data, can be found in Ref. \cite{Re99}.

\section{Conclusions}

We suggested a simple parametrization of the three deuteron electromagnetic FFs, with a minimal number of parameters, based on a transparent physical picture. It can be used in the comparison of different theoretical models with  experiments involving deuterons, and for a precise analytical interpolation of the experimental points in the region $Q^2< 2$ GeV$^2$. The present parametrization is based on a classical description of the deuteron structure, in terms of hadronic (nucleon+meson) degrees of freedom. The region of $Q^2$, where the separation of the charge and quadrupole deuteron FFs has been done, can not be easily  extended in next future \cite{Gi03}. For $Q^2\ge 1.8$ GeV$^2$,  polarization measurements (which normally require a secondary scattering or a polarized target) are extremely challenging at the present accelerators, with the present techniques of polarized targets or polarimeters due to the steep decreasing of the cross section. 

The parametrization proposed here can be considered a generalization of the model \cite{Ia73}, developped for the nucleon electromagnatic FFs. As in the nucleon case, the considered parametrization obeys, by construction, to the analyticity properties of FFs and can be extended to the time-like region. This is the object of a forthcoming work.

\section{Acknowledgments}
The authors acknowledge Prof. M. P. Rekalo for enlightning discussions and ideas, without which this paper would not have been realized in the present form.

We thank Prof. F. Iachello for his interest to this work and for useful comments.

{}

\clearpage\newpage

\clearpage\newpage

\clearpage\newpage
\clearpage\newpage
\clearpage\newpage
\clearpage\newpage
\end{document}